\documentstyle[preprint,pra,aps]{revtex}
\tightenlines
\begin{document}
\draft
\title{Space-Time Variation of Physical Constants and Relativistic 
Corrections in Atoms}

\author{ V. A. Dzuba$^*$, V. V. Flambaum, J. K. Webb }
\address{School of Physics, University of New South Wales,
	Sydney, 2052, Australia}
\date{\today}
\maketitle
%------------------------------------------------------------------
\begin{abstract}
Detection of high-redshift absorption in the optical spectra of quasars
have provided a powerful tool to measure spatial and temporal variations
of physical ``constants'' in the Universe. It is demonstrated that high 
sensitivity to the variation of the fine structure constant $\alpha$ can be 
obtained from a comparison of the spectra of heavy and light atoms
(or molecules). We have performed calculations for the pair FeII and MgII
for which accurate quasar and laboratory spectra are available.
A possibility of $10^5$ times enhanced effects of the fundamental
constants variation suitable for laboratory measurements is also discussed.
\end{abstract}

\pacs{31.30.Jv , 95.30.Dr , 95.30.-k}

%------------------------------------------------------------------

The first ideas about possible variation of the fundamental physical constants
in the expanding Universe were suggested by Dirac \cite{Dirac}. Now this
subject is of particular current interest because of the new possibilities 
opened up by the structure of unified theories, like the string theory and
M-theory, which lead us to expect that additional compact dimensions of
space may exist. The ``constants'' seen in the three-dimensional subspace of
the theory will vary at the same rate as any change occurring in the extra 
compact dimensions (see, e.g. \cite{Marciano84,Barrow87,Damour94}).

Quasar absorption systems present ideal laboratories where to search
for any temporal or spatial variation of fundamental constants by comparing
atomic spectra from the distant objects with the laboratory spectra
(see, e.g. \cite{Potechin95} and references therein).

The energy scale of the atomic spectra is given by the atomic  unit 
$\frac{me^4}{\hbar^2}$. In the absence of any corrections, all atomic
spectra are proportional to this constant and no change of the fundamental
constants can be detected. Indeed, any change in the atomic unit will be 
absorbed in the determination of the red shift parameter 
$1 + z = \frac{\omega}{\omega '}$ which is also found from the
comparison between the cosmic and laboratory
 atomic spectra ($\omega '$ is the red-shifted frequency of
the atomic transition and $\omega$ is this frequency in the laboratory).
However, any change of the fundamental constants can be found by measuring the
 relative size of relativistic corrections, which are proportional
to $\alpha^2$, where $\alpha = e^2/\hbar c$ is the fine structure constant
\cite{other}.

It would seem natural to find this change from  measurements of the
 spin-orbit
splitting within a fine-structure multiplet. However, this way is not the
most efficient, and it may even give incorrect results, since other
relativistic effects are ignored. 
The aim of this letter is to demonstrate that the change in $\alpha$
produces  an order of magnitude
larger effect in the difference between
transition frequencies in heavy and light atoms (or molecules). We have
calculated the dependence
of the transition frequencies on $\alpha$ for FeII (see eq.(\ref{Fe}))
and MgII (see eq.(\ref{Mg})) where accurate data exist both for laboratory
and quasar spectra. Other possibilities include
 comparisons of different optical transitions, for example $s-p$ and $p-d$,
 in the same atom or molecule,
or  comparisons of microwave transitions in molecules which contain rotational
and hyperfine intervals.

 We also propose another interesting possibility: to use transitions between
``accidentally'' degenerate levels in the same atom or molecule. This
degeneracy would disappear after a minor change in $\alpha$.
For example, in the Dy atom  there are two degenerate opposite-parity levels
 \cite{DzubaDy,Budker}. 
The frequency of the $E1$-transition between them is smaller then the 
hyperfine splitting of each level. As a result,
 the relative effect of the change in $\alpha$ is enhanced by five orders
of magnitude (the ratio of the size of the relativistic effect to the
 transition frequency). This case seems to be
more  suitable for laboratory experiments. Similar experiments with 
``accidentally'' degenerate  molecular levels belonging to different electron
terms are also sensitive to the smallest changes in the ratio of the nucleon 
to electron masses since in this case the difference in electron energies 
is compensated by the difference in the vibrational and rotational energies 
of the nuclei. As is known, the nuclear mass is a function of the strong 
interaction constants and vacuum condensates.

A competitive possibility is an accurate measurement of a very small
difference between the frequencies of two transitions in different atoms 
or molecules (like in the comparison with the frequency standard). 
This small difference can be measured with a very high absolute accuracy,
up to few Hz, and one can have the same enhanced effects of the change 
in $\alpha$ or the nucleon mass.  

Let us start our calculations from simple analytical estimates of the
relativistic effects in transition frequencies.
Consider first the relativistic corrections to the frequency of an atomic 
transition in a hydrogen-like atom. The relativistic correction to the energy
level is given by (see, e.g. \cite{Drell})
\begin{eqnarray}
	\Delta_n = - \frac{me^4Z^2}{2\hbar^2}\frac{(Z\alpha)^2}{n^3}
	(\frac{1}{j + 1/2} - \frac{3}{4n}),
\label{rel}
\end{eqnarray}
where $Z$ is the nuclear charge, $n$ is the principal quantum number and $j$ is
the total electron angular momentum. This value of the relativistic correction 
can be obtained as an expectation  value $\langle V \rangle$ of the 
relativistic perturbation $V$, which is large in the vicinity of the 
nucleus only. Therefore, the relativistic correction
$\Delta$ is proportional to the electron density near the nucleus
 $|\Psi(r<\frac{a}{Z})|^2 \propto \frac{Z^3}{n^3 a^3}$
($a$ is the Bohr radius, $\frac{a}{Z}$ is the size of the hydrogen-like ion).
 For an external electron in a many-electron
atom or ion the electron density near the nucleus is given
by the formula (see, e.g. \cite{Sobelman}) obtained in the semiclassical
approximation ($n \gg 1$)
\[
	|\Psi(r < \frac{a}{Z})|^2 \propto \frac{Z^2_a Z}{\nu^3 a^3},
\]
where $Z_a$ is the charge ``seen'' by the external electron outside the
atom, i.e. $Z_a$  = 1 for neutral atoms, $Z_a$ = 2 for singly charged ions,
etc.; $\nu$ is the effective principal quantum number, defined by 
$E_n = - \frac{me^4}{2\hbar^2}\frac{Z_a^2}{\nu^2}$, where $E_n$ is 
the energy of the electron. For hydrogen-like ions $\nu = n, Z_a = Z$.
Thus, to find the single-particle relativistic correction, we should
multiply $\Delta$ in Eq. (\ref{rel}) by the ratio of $|\Psi(r<\frac{a}{Z})|^2$
in the multi-electron ion and hydrogen-like ion. The result is
\begin{eqnarray}
	\Delta_n = - \frac{me^4 Z_a^2}{2\hbar^2}\frac{(Z\alpha)^2}
	{\nu^3} \left[\frac{1}{j+1/2}-\frac{Z_a}{Z\nu}(1-\frac{Z_a}{4Z})\right]
 \simeq E_n \frac{(Z\alpha)^2}{\nu (j+1/2)}.
\label{rel1}
\end{eqnarray}
The second term in the square brackets is presented  to provide a
  continuous transition from the hydrogen-like ion  Eq. (\ref{rel}) to
the multi-electron ion  Eq. (\ref{rel1}). In multi-electron ions  ($ Z \gg 
Z_a$) this term is, in fact, a rough  estimate based on the direct calculation 
of $\langle V \rangle$.  We should neglect this small term  since there
are more important many-body corrections
(the accurate many-body calculations discussed below give approximately
constant correction term in square brackets with the value about $-0.6$).

We see that the relativistic correction is largest for the $s_{1/2}$ and 
$p_{1/2}$ states, where $j = 1/2$. The fine structure splitting is given by
\begin{eqnarray}
\Delta_{ls} = E(p_{3/2}) - E(p_{1/2}) \simeq - \Delta(p_{1/2})/2 \simeq 
 - \Delta(p_{3/2}).
\label{dls}
\end{eqnarray}

In the quasar absorption spectra transitions from the ground state have 
been observed. Therefore, it is important to understand how the frequencies
of these transitions are affected by the relativistic effects.
The fine splitting in excited states is smaller then the
relativistic correction in the ground state, since the density of
the excited electron near the nucleus is smaller. As a result, the fine 
splitting of the $E1$-transition from the ground state (e.g., $s -p$) is
substantially smaller than the absolute shift of the frequency of the 
$s -p$ transition. The mean energy of the $p$-electron is defined as
\[
	E(p) = \frac{2}{3}E(p_{3/2}) + \frac{1}{3}E(p_{1/2}) \simeq
	E_n(p) - \frac{4}{3}\Delta_{ls},
\]
where $E_n$ is the non-relativistic energy. Therefore, the relativistic shift
of the mean $s -p$ transition frequency is given by
\[
	\Delta(p-s) \simeq - \frac{4}{3}\Delta_{ls} - \Delta(s_{1/2}).
\]
The relative size of the relativistic corrections is proportional to
 $Z^2$, so they are small in
light atoms. Therefore, we can find the change of $\alpha$ by comparing 
transition frequencies in heavy and light atoms. We stress that the
most accurate and effective procedure 
to search for the change of $\alpha$ must include all 
 relativistic corrections and the analysis of 
all available lines (rather then the fine splitting within one
 multiplet only).

Let us consider an example of the $s -p$ transitions from the ground state
in MgII (Z=12) and FeII (Z=26) ions where accurate data exist for both
 laboratory and 
quasar spectra (see detailed analysis in \cite{Webb,Mg,Fe}).
MgII is a simple system with one external electron above closed shells.
The frequency of the transition can be presented in the following form
\begin{equation}
\begin{array}{ll}
    E(3p_{1/2}) - E(3s_{1/2}) = & E_l(3p_{1/2}) - E_l(3s_{1/2}) +
    K_1 [(\frac{\alpha}{\alpha_l})^2 -1] \\
    E(3p_{3/2}) - E(3s_{1/2}) = & E_l(3p_{3/2}) - E_l(3s_{1/2}) +
    K_2 [(\frac{\alpha}{\alpha_l})^2 -1] \\
\end{array}
\label{EE}
\end{equation}
\begin{equation}
\begin{array}{ll}
    K_1 \simeq & -2\Delta_{ls} - \Delta(3s_{1/2}) \, , \,\,\,    \\
 K_2 \simeq & -\Delta_{ls} - \Delta(3s_{1/2}).\\
\end{array}
\label{K1}
\end{equation}
Here $\alpha_l$ and $E_l$ are the laboratory values of fine structure constant
and energy, $\alpha$ and $E$ are the values at the distant object. 
The formulae (\ref{rel1})-(\ref{dls}) for the spin-orbit splitting 
$\Delta_{ls}$ and relativistic shift $\Delta(3s_{1/2})$ have been 
obtained in the single-particle  approximation.
There are large  corrections to these formulae  due to partial electron 
screening of the nuclear potential, which appears in the equation for the
relativistic correction $\langle V \rangle$, and due to a change of the total 
atomic potential acting on external electron when we change $\alpha$.
Indeed, the wave functions of all inner electrons change due to the
 relativistic 
corrections ($\Psi \rightarrow \Psi + \delta \Psi$), which in turn
 produces a
change in the mean-field atomic potential ($V \rightarrow V + \delta V$).
In part these effects can be taken into account by using the experimental
value of $\Delta_{ls}$ = 91.6 cm$^{-1}$ and the semiempirical formula for
$\Delta(3s) = -2 \Delta_{ls}(\frac{\nu_{3p}}{\nu_{3s}})^3 =-3.37\Delta_{ls}=
-309$  cm$^{-1}$ obtained from Eqs. (\ref{rel1})-(\ref{dls}).
This yields $K_1 = 126$  cm$^{-1}$ and  $K_2 = 217.6$ cm$^{-1}$.
To obtain more accurate values of $K_1$ and $K_2$ we performed
calculations of the MgII spectra using many-body perturbation theory.
We have used the complete set of the  relativistic Hartree-Fock
energies and wave functions as a zero approximation and then calculated
 all second-order correlation
corrections in the residual electron-electron interaction (this technique is
described in \cite{Dzuba}). 
This {\it ab initio} calculation reproduces the experimental
energy levels of the external electron with  $0.2\%$ accuracy (the
single-electron energy levels in the many-body problem are defined as the
ionization energies with a minus sign).
 To find the value of relativistic
corrections we performed calculations for the two values of $\alpha:
 \alpha = \alpha_l$ and $\alpha = \alpha_l/2$. The calculated relativistic
corrections to the energy levels  of the external electron are :$\Delta(3s) =
-189.4 ,\, \Delta(3p_{1/2}) =-72.1 , \, \Delta(3p_{3/2})= 24.4 , \,
 \Delta_{ls}= 96.5  $cm$^{-1}$. Note
that the many-body corrections change the sign of $ \Delta(3p_{3/2})$.
 The final results are the following:
\begin{eqnarray}
\begin{array}{ll}
	3s_{1/2} - 3p_{1/2}: & \omega_1 = 35669.26 + 119.6 
	[(\frac{\alpha}{\alpha_l})^2 - 1] \,\, \mbox{cm}^{-1} \\
	3s_{1/2} - 3p_{3/2}: & \omega_2 = 35760.834(4) + 211.2 
	[(\frac{\alpha}{\alpha_l})^2 - 1] \,\, \mbox{cm}^{-1} \\
\end{array}
\label{Mg}
\end{eqnarray}
We changed slightly the values of the calculated coefficients $K_1$ and $K_2$
to reproduce exactly the experimental value of the fine
 splitting $\Delta_{ls}= 91.6$cm$^{-1}$. 
The laboratory frequency of the $3s_{1/2} - 3p_{3/2} $ transition in the
main isotope $^{24}$Mg (abundance 79 \%) has been taken from \cite{Mg}. The
increase of the frequencies (isotope shifts) for the transition in
 $^{25}Mg$ (abundance 10 \%) and $^{26}$Mg (abundance 11 \%) are equal 
to 0.053 and 0.102 $cm^{-1}$, respectively \cite{Mg}.
The laboratory frequency of the $3s_{1/2} - 3p_{1/2} $ transition has
been taken from \cite{Webb}.

Now consider FeII. We are interested in the $E1$-transition from the ground 
state 
$3d^6 4s ~~^6D_{9/2}$ to the members of the multiplets $3d^6 4p ~~^6D, 
 ~~^6F$ and ~~$^6P$. Due to the selection rule $ \Delta J \leq 1 $ the
transitions from the ground state $J=9/2$ can involve only few components 
of the excited multiplets. 
The splitting between the nearby components of the same multiplet
 (e.g. $J$ = 9/2 
and $J$ = 7/2) are about 100 - 200 cm$^{-1}$. On the other hand,
the total relativistic correction 
to the mean frequency of the $4s - 4p$ transition is about 1800 cm$^{-1}$, i.e.
 $\sim$ 10  times larger! Unfortunately, we can not measure the effect of this
large correction by measuring the frequency of  $4s - 4p$ 
transitions in one element (FeII) since it will be hidden in the definition
 of the red shift 
parameter $1 + z$. Therefore, to make use of this large effect we need to
consider two elements with different relativistic shifts, e.g. MgII and FeII,
or different transitions in the same element, e.g. $s -p$ and $d - p$.
 
   We can also consider $s-p$ transitions to the different multiplets. However,
the relativistic effects in the differences between the central energy points
$E_0$ of different multiplets ( $D , F$ and $P$)
are substantially smaller ( $\sim$ 100 - 200 cm$^{-1}$) than the relativistic
 shift of the  $s-p$ transition frequency, since these 
differences are due to the dependence of the Coulomb interaction $Q$ 
between external electrons on the relativistic correction to their wave
functions ($\Delta \omega \sim Q Z^2 \alpha^2$ where $Q$ is the non-diagonal
matrix element of the Coulomb interaction producing configuration mixing).
There are also small relativistic corrections to the interval between the
 different
multiplets due to the second order in the spin-orbit interaction. However, the 
relativistic effect in the difference of the frequencies of particular
components of different multiplets can be  larger. The maximal
relativistic shift in the differences of the FeII
 frequencies (for the transitions which have been observed in the quasar
 spectra)  is  387 cm$^{-1}$ in the difference between 
$^6F_{9/2}$ and $^6P_{7/2}$ (see below).

In principle, approximate calculations of the average relativistic 
frequency shift of the $4s - 4p$ transition and the dependence of the
fine structure intervals on $\alpha$ could be done using semiempirical 
formulae (\ref{rel1}) - (\ref{K1}) which give average the shift 
1550 cm$^{-1}$, and laboratory experimental data. However, to calculate
accurately the relativistic shift for each transition we performed relativistic
many-body calculations for energy levels of FeII.
\begin{enumerate}
\item We used averaged relativistic Hartree-Fock potential of the
FeIII $d^6$ state to generate a complete set of the zeroth approximation 
wave functions and energies.
\item Correlations of the second-order in the residual Coulomb interaction 
between the valence and core electrons were included by means of the 
correlation potential (self - energy operator) method \cite{Dzuba}.
\item Most of the valence correlations were included by means of a
configuration interaction method. Excited configurations were constructed 
from the $s, p$ and $d$ single-electron Hartree-Fock states with $n \leq 6$
(17 single-electron orbitals). 
Only those configurations were included that can be obtained by a single
 or double-electron excitation from the main configuration.
 Thus, the total number of configurations was
few hundred. This is a rather small-scale configuration interaction and 
full convergence was not achieved. However, good agreement with experimental
data for the energy levels shows that a greater part of the valence
correlations were included.
\item To imitate the effect of higher-order many-body corrections we 
introduced two fitting parameters. The {\it ab initio} correlation potentials 
for the $s$ and $p$ electrons were multiplied by the factors 
$f_s = 0.94$ and $f_p =0.9$, respectively.
These factors imitated the effect of screening of the Coulomb 
interaction between the core and valence electrons which  reduces 
the correlations between these electrons (accurate calculation of this
screening has been done in \cite{Dzuba}). 
The values of $f_s$ and $f_p$ were chosen to fit the ionization energies
of the $3d^6 4s$ and $3d^6 4p$ states.
\end{enumerate}
To find the dependence of frequencies on $\alpha$ we use the following 
formula for the energy levels within one fine-structure multiplet:
\begin{eqnarray}
	E = E_0 + Q_1 ((\frac{\alpha}{\alpha_l})^2-1) +
	K_1 ({\bf LS})(\frac{\alpha}{\alpha_l})^2 +
	K_2 ({\bf LS})^2 (\frac{\alpha}{\alpha_l})^4.
\label{ls}
\end{eqnarray}
Here $L$ is the total orbital angular momentum and $S$ is the total
electron spin.
We introduce an $({\bf LS})^2$ term to describe deviations from the 
Lande interval rule. There are two sources of the $({\bf LS})^2$ term:
the second order in the spin-orbit interaction ($ \sim (Z \alpha)^4 = 
1.3 \times 10^{-3}$) and the first order in the spin-spin interaction 
($ \sim \alpha^2 = 5.3 
\times 10^{-5}$). In FeII the second-order spin-orbit interaction is larger.
Therefore, we first fitted the experimental fine structure intervals to find
$K_1$ and $K_2$ (numerical calculations give close values of $K_1$ and $K_2$).
Then we used numerical calculations for $\alpha = \alpha_l$ and
$\alpha = \alpha_l/2$ to find the dependence of the configuration centers
$E_0$ on $\alpha$ (coefficient $Q_1$). The coefficients $Q_1, K_1$ and $K_2$
are presented in table I. Very accurate values of the laboratory
frequencies ($\alpha = \alpha_l$) can be 
found in \cite{Fe}. The errors are about 0.003 cm$^{-1}$. The Fe atom
 has one dominating isotope $^{56}$Fe (92 \%) and small isotope shifts. 
 Now we can use Eq. (\ref{ls}) and table I  to calculate
the  frequencies of the $E1$ transitions from 
the ground state (in cm$^{-1}$) as functions of $\alpha$:\begin{equation}
\begin{array}{ccc}
 ^6 D & J= 9/2 & \omega = 38458.9871 + 1394x + 38y , \\
      & J= 7/2 & \omega = 38660.0494 + 1632x + 0y  , \\
 ^6 F & J=11/2 & \omega = 41968.0642 + 1622x + 3y , \\
      & J= 9/2 & \omega = 42114.8329 + 1772x + 0y , \\
      & J= 7/2 & \omega = 42237.0500 + 1894x + 0y ,  \\
 ^6 P & J= 7/2 & \omega = 42658.2404 + 1398x - 13y ,  \\
\end{array}
\label{Fe}
\end{equation}
where
$x = (\frac{\alpha}{\alpha_l})^2 - 1, y = (\frac{\alpha}{\alpha_l})^4 - 1$.
One can use Eqs. (\ref{Mg}) and (\ref{Fe}) to fit FeII and MgII lines in the
quasar
spectra and find the variation of $\alpha$. Note that besides FeII one can
use $s - p$ transitions from the ground state of GeII, ZnII, NiI, FeI,
MnII, CaI and CaII, where the relativistic corrections have the same order of 
magnitude, and any light atom besides MgII.

One more interesting possibility more suitable for a laboratory
experiment is to use transitions between ``accidentally'' degenerate levels 
in the same atom. Such meta-stable levels exist, for example, in the Dy atom:
two $J=10$ opposite parity levels $4f^{10}5d6s$ and $4f^95d^26s$ lying 19797.96
cm$^{-1}$ above ground state.( This pair of levels was used to study
parity non-conservation in Refs. \cite{DzubaDy,Budker}). There are other
 examples of ``accidentally'' degenerate levels in the rare-earth
and actinide atoms
and many close levels in other heavy atoms and ions (in the absence of
 degeneracy one should look for $s-d$ or $s-p$ transitions where the 
relativistic effects are larger).
In the case of ``accidental'' degeneracy the contributions of the
 relativistic corrections to the frequency
of the $E1$ transition in a heavy atom ($\sim 1000 cm^{-1}$) is compensated
by the difference in the Coulomb interaction energies of the two 
configurations. 
However, if $\alpha$ varies with time, this compensation will eventually 
disappear. Thus, we have a correction 
$\sim 1000$ cm$^{-1} ((\frac{\alpha}{\alpha_l})^2 -1)$ to the very small
( < 0.01 cm$^{-1}$) frequency of the transition. One can measure,
for example, the time dependence of the ratio of frequencies for
transitions between the hyperfine components of these two states.
In the case of ``accidentally'' degenerate levels belonging to  
different electron terms in a molecule one can have enhanced effects 
of the change of both $\alpha$ and the nucleon mass. 
In the latter case the enhancement
factor is the ratio of the vibration energy to the small frequency of the
 transition. 

We are grateful to O. Sushkov and D. Budker for useful discussions.

%####################################################################

\begin{table}
\caption{Relativistic energy parameters of some FeII multiplets (cm$^{-1}$)
%(Eq. (\ref{ls}))}
(Eq. (7))}
\label{tab}
\begin{tabular}{ccccc}
Multiplets & $E_0$ & $K_1$ & $K_2$ & $Q_1$~~ \\
\hline
 $~~^6D$ & 38686.19 & -53.034 & 1.5189 & 1659~~ \\
 $~~^6F$ & 42168.91 & -27.136 & 0.04741 & 1826~~ \\
 $~~^6P$ & 43078.15 & -162.602 & -2.145 & 1805~~ \\
\end{tabular}
\end{table}
\end{document}